Giovanni Cherubin*

# Bayes, not Naïve: Security Bounds on Website Fingerprinting Defenses

**Abstract:** Website Fingerprinting (WF) attacks raise major concerns about users' privacy. They employ Machine Learning (ML) techniques to allow a local passive adversary to uncover the Web browsing behavior of a user, even if she browses through an encrypted tunnel (e.g. Tor, VPN). Numerous defenses have been proposed in the past; however, it is typically difficult to have formal guarantees on their security, which is most often evaluated empirically against state-of-the-art attacks. In this paper, we present a practical method to derive security bounds for any WF defense, where the bounds depend on a chosen feature set. This result derives from reducing WF attacks to an ML classification task, where we can determine the smallest achievable error (the Bayes error). Such error can be estimated in practice, and is a lower bound for a WF adversary, for any classification algorithm he may use. Our work has two main consequences: i) it allows determining the security of WF defenses, in a black-box manner, with respect to the state-of-the-art feature set and ii) it favors shifting the focus of future WF research to identifying optimal feature sets. The generality of this approach further suggests that the method could be used to define security bounds for other ML-based attacks.

## 1 Introduction

Secure network protocols often fail at hiding the length of packets, their timing, or the total communication bandwidth. In the past, it was shown that Traffic Analysis (TA) adversaries can use such information to profile users, break the confidentiality of their communications, or violate their anonymity [2, 11, 29].

*Corresponding Author: Giovanni Cherubin: Royal Holloway University of London, UK, E-mail: Giovanni.Cherubin.2013@live.rhul.ac.uk

Website Fingerprinting (WF) attacks are an important subclass of TA. In WF, a local adversary observes the encrypted network traffic produced by a victim loading a web page, selects high level descriptions of this traffic (*features*), and uses a Machine Learning (ML) classifier on them to predict which page the victim visits, even if she browses through an encrypted tunnel (e.g., Tor, VPN). In the specific case of Tor, it was shown that a WF adversary controlling the Guard node of a victim browsing an .onion site can perform this attack with high success probability [16]. Since the attack would go undetected, and because of the harm such anonymity breach can cause to the victim, WF attacks constitute a serious threat for the Tor network.

WF defenses have been proposed in the past. Unfortunately, the most common method to evaluate their security so far has been testing them against state-of-the-art attacks. This inevitably created an arms race, which has continued for more than a decade [5, 11, 13, 17, 18, 20, 26, 30]. The concerns raised by novel attacks led researchers to reconsider defense BuFLO [11], which was initially proposed as a proof-of-concept representing very secure but inefficient countermeasures, as a candidate for real-world applications [6].

The first study towards provable security evaluation of WF defenses considered an idealized lookup-table adversary, who knows the exact network traffic corresponding to each web page, and only commits an error when many pages produce identical traffic [5]. Unfortunately, this method is highly susceptible to noise, as a small difference in the measured quantities strongly impacts the results. Furthermore, this approach is not applicable to a large class of defenses (probabilistic defenses), which is a dramatic constraint for defenses designers. Because of these shortcomings, most defenses are still evaluated against state-of-the-art attacks.

In this paper, we provide a natural formalization of WF attacks as an ML classification problem. This allows showing that the success of a WF adversary is bounded, with respect to the features he uses, by the success of the ideal Bayes classifier. We then indicate a lower bound estimate of the Bayes error, which is asymptotically the smallest error achievable by any classifier.

We summarize our contributions as follows.

| Defense | Closed World | | One VS All | | Overhead | |
|---|---|---|---|---|---|---|
| | $\hat{R}^*$ (%) | $(\varepsilon, \Phi)$-*privacy* | $\hat{R}^*$ (%) | $(\varepsilon, \Phi)$-*privacy* | Packet (%) | Time (%) |
| No Defense | 6.2 ± 0.3 | (0.06, k-NN) | 2.5 ± 1.7 | (0.05, k-NN) | 0.0 | 0.0 |
| Decoy Pages | 42.6 ± 0.6 | (0.43, k-NN) | **14.7 ± 5.5** | **(0.29, k-NN)** | 134.4 | 59.0 |
| BuFLO | 56.9 ± 1.0 | (0.58, k-FP) | 14.4 ± 13.7 | **(0.29, k-FP)** | 110.1 | 79.1 |
| Tamaraw | **69.0 ± 0.9** | **(0.70, k-NN)** | 12.7 ± 5.8 | (0.25, k-NN) | 257.6 | 341.4 |
| CS-BuFLO | 61.9 ± 0.9 | (0.63, k-FP) | 8.1 ± 3.5 | (0.16, k-FP) | 67.2 | 575.6 |
| WTF-PAD | 48.6 ± 1.2 | (0.49, k-FP) | 9.0 ± 3.8 | (0.18, CUMUL) | 247.0 | 0.0 |

**Table 1.** Bayes lower bounds $\hat{R}^*$ and $(\varepsilon, \Phi)$-*privacy* of WF defenses on the WCN+ dataset. A lower bound $\hat{R}^*$ is the smallest error an adversary achieves in a specific scenario (Closed World, One VS All). The metric $(\varepsilon, \Phi)$-*privacy* indicates how far a defense is from perfection (achieved for $\varepsilon = 1$). In bold, the most $(\varepsilon, \Phi)$-*private* defenses. We use the name of an attack to indicate its feature set $\Phi$.

*Security bounds*

We show that a WF adversary using a certain feature set is bounded by the Bayes error, of which we can estimate a lower bound. This is guaranteed to be an asymptotic lower bound of the adversary's error, for any classifier he may choose. The bound holds for arbitrary distributions of data, under the standard i.i.d. assumption. We verify the bound in practice, for a finite number of examples.

*$(\varepsilon, \Phi)$-privacy*

We propose a privacy metric, $(\varepsilon, \Phi)$-*privacy*, based on the lower bound error estimate. The metric indicates how far a defense is from a perfectly private one ($\varepsilon = 1$), and depends on a feature set $\Phi$. It can be used to evaluate the security of any defense, in a black-box manner. This result also allows WF researchers to shift their attention to identifying optimal features.

*Evaluation*

We compare the most influential defenses in terms of their $(\varepsilon, \Phi)$-*privacy* and overheads (Table 1). Furthermore, we use the lower bound estimates to evaluate feature sets, and notice that they have not improved substantially in the past 3 years. This suggests it may be difficult to reduce significantly the bounds we computed with the current state-of-the-art feature sets.

*Code*

We wrote a comprehensive Python framework to perform tests on WF attacks and defenses, which collects the efforts of WF researchers into a unified API. We also implemented our method to derive security bounds on a defense. Both can be found at:

https://github.com/gchers/wfes.

## 2 Website Fingerprinting

We presently provide an informal description of WF attacks; we will formalize these concepts in section 4.

In WF attacks, a passive eavesdropper observes the encrypted traffic produced by a victim while she loads web pages via an encrypted tunnel (e.g., a VPN, Tor). It is assumed that the adversary does not know the IP address of the web server to which the victim sends her requests, and that he cannot decrypt the packets. However, he is generally allowed to know which defense the victim uses. The adversary selects a set of web pages (*monitored* pages), and his goal is to predict which of those pages the victim visits, if any.

A *Closed World* scenario is when the victim is only allowed to visit one of the monitored pages (Figure 1a). This scenario is a simplification of real-world attacks, but because it allows an adversary to perform the attack in nearly-perfect conditions, it is used for evaluating the security of defenses. An *Open World* scenario resembles more closely a real-world setting, where the victim can visit both monitored and unmonitored pages. This scenario is typically used for evaluating attacks with respect to their scalability to the real world. In addition to the Closed World scenario, we will evaluate defenses under the One VS All scenario, a special case of Open World, which indicates the ability of a defense to protect individual pages (Figure 1b). In the One VS All scenario the adversary only monitors one web page, and the victim can either visit that page, or one of the *unmonitored* pages; for this reason, this scenario can be seen as an extreme case of the Open World scenario.

A WF attack has two phases: *training* and *attack*. In the training phase, he generates defended network traffic on his computer by loading monitored pages; then, he extracts features from this traffic, and trains an ML classifier on them. Features are high level descriptions

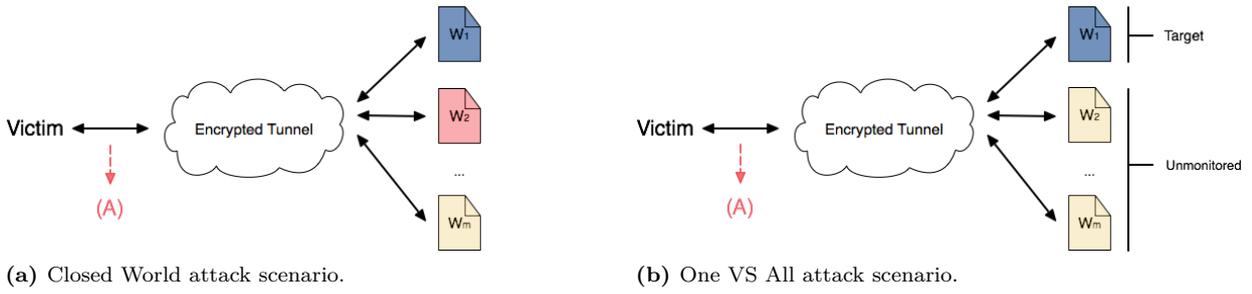

**(a)** Closed World attack scenario.

**(b)** One VS All attack scenario.

**Fig. 1.** In the Closed World scenario (a), the victim visits one page among those monitored by the adversary. In Open World, the victim is allowed to visit unmonitored pages. In One VS All (b), a special case of Open World, the adversary only monitors one page.

of data (network traffic, in this case), and will be formalized in section 4; our work demonstrates that the choice of features crucially impacts the success of WF attacks. The training phase can be performed offline. In the attack phase, the adversary collects defended network traffic coming from the victim. He extracts features from the new traffic, and uses the ML classifier to predict which web pages the victim loads.

## 3 Related Work

In this section, we will describe the major WF attacks and defenses, and previous directions in provable evaluation of defenses.

### 3.1 Major defenses

WF defenses can be divided into *deterministic*, which deterministically transform the characteristics of traffic, and *probabilistic* (also called elsewhere *random*), where randomness is involved in the morphing process. We will consider six defenses: two deterministic (BuFLO and Tamaraw), four probabilistic (Randomized Pipelining, Decoy Pages, CS-BuFLO, WTF-PAD).

Designing a WF defense requires: i) understanding of how WF attacks work and what they target, and ii) engineering the defense on the top of other network protocols. In the past, probabilistic defenses have attracted the attention of researchers because they are often easier to implement, and they tend to introduce less overhead.

*RP*
Randomized Pipelining (RP) randomizes the order of HTTP requests in the HTTP pipeline. This defense is embedded by default in the Tor browser. Since in experiments we will defend network traffic coming from the Tor browser, we will implicitly assume RP is underlying to any of the defenses we shall now present; consequently, "No Defense" will refer to plain Tor traffic (i.e., only defended using RP) [21].

*BuFLO*
For parameters $(d, \rho, \tau)$, BuFLO sends packets of fixed size $d$, with frequency $\rho$, for at least $\tau$ time. If more time is required by the page load, packets of size $d$ are sent with frequency $\rho$ for the time needed [11].

*Tamaraw*
Similarly to BuFLO, it sends fixed-size packets at fixed intervals. It considers distinct intervals between packets with respect to their direction: outgoing packets are sent every $\rho_{out}$ seconds, incoming packets are sent every $\rho_{in}$ seconds, and $\rho_{out} > \rho_{in}$. It further pads the number of packets sent in both directions by multiples of a padding parameter [5].

*Decoy Pages*
Loads a randomly chosen page in background together with the requested page. The traffic introduced by the background page should confuse features and classifier used by the adversary. It is a probabilistic defense, and it is arguably the easiest to implement among the defenses presented here [20].

*CS-BuFLO*
CS-BuFLO is a modification of BuFLO, which reduces overheads and eases implementation. As with BuFLO, it sends packets of size $d$ with frequency $\rho$. However, the value of $\rho$ is dynamically adapted to the network bandwidth, and events like client's onLoad (i.e., the browser has finished loading a page) and channel_idle are used to determine the end of communication. Padding is also inserted at the end of communication, up to a certain transmission size [4].

*WTF-PAD*

WTF-PAD is based on Adaptive Padding (AP) [23]. At each endpoint (client, server), the defense sends real or dummy packets according to the state (idle, burst, gap) of an internal state machine. This allows adaptively morphing traffic to avoid large overheads [15]. This defense is relatively easy to implement, although it is currently not well understood how to optimally determine the distributions that regulate transitions of the internal state machines. For experiments we used the distributions proposed by Pulls [22].

## 3.2 Major attacks

We now itemize the most influential WF attacks. A WF attack is denoted by a feature set and an ML classifier. This will be formalized in section 4.

*LL*

The feature set of this attack is composed of the count of packets with a certain direction and size, for each direction and size $\{\uparrow, \downarrow\} \times (0, ..., MTU)$, where $MTU = 1500$ is the maximum transmission unit. It uses the naïve Bayes classifier (NB) for classification, with kernel density estimation (KDE) for estimating the conditional probabilities [17].

*VNG++*

Its feature set includes total time span of a trace, total per-direction bandwidth, and *(direction, size)* of each sequence of contiguous outgoing packets ("bursts"). It uses NB as a classifier [11].

*k-NN*

The feature set comprises general features (e.g., bandwidth, and packet counts), unique packet lengths, features related to packet ordering and bursts, and the lengths of the first 20 packets accounting for direction (set to negative for incoming packets). An algorithm also determines a set of weights for the features, according to their importance; feature vectors are multiplied by these weights before classification[1]. For classification, it uses a custom modification of the k-Nearest Neighbors classifier, which works as follows. To predict the label for an object $x$, the classifier first determines the $k$ closest objects to $x$. If all of them have the same label, it predicts that label; otherwise, it outputs an empty prediction. Manhattan distance is used as a distance metric for the classifier [26].

| Rank | Feature Description |
|---|---|
| 1 | # incoming packets |
| 2 | # outgoing packets (fraction of tot # packets) |
| 3 | # incoming packets (fraction of tot # packets) |
| 4 | Standard deviation of outgoing packet ordering list |
| 5 | # outgoing packets |
| 6 | Sum of items in the alternative concentration feature list |
| 7 | Average of outgoing packet ordering list |
| 8 | Sum of incoming, outgoing, and total # of packets |
| 9 | Sum of alternative # packets per second |
| 10 | Tot # packets |
| 11-18 | Packet concentration and ordering feature list |
| 19 | # incoming packets stats in first 30 packets |
| 20 | # outgoing packets stats in first 30 packets |

**Table 2.** Best 20 features according to the recent feature analysis by Hayes and Danezis [12].

*CUMUL*

The feature set of this attack includes basic information of packet sequences (e.g., total bandwidth, total in/out packets), together with the cumulative sum of packets' sizes. The attack employs a Support Vector Machine (SVM) classifier with an RBF kernel, and uses cross validation (CV) grid search to determine the optimal parameters for the kernel [19].

*k-FP*

This is currently the state-of-the-art attack. The feature set of this attack comes from a systematic analysis of the features proposed by previous research. Importance of features was evaluated with respect to the classifier used for the attack (k-Nearest Neighbors); the 20 most important features are in Table 2. In this attack, the selected features are transformed using Random Forest (RF) as follows. First, feature values from the original feature set are extracted. Then, RF is applied to these values for generating leaves; leaves are then used as feature values for classification. As a classifier, this attack employs the custom modification of the k-Nearest Neighbors classifier that was used for the k-NN attack. Hamming distance is used as a distance metric [12].

---

[1] In the original attack, weights are used by the distance metric. This is equivalent to multiplying weights before prediction.

### 3.3 Previous directions in provable evaluation of WF defenses

Cai et al. [5] proposed a method to compute a lower bound of error for WF adversaries. They considered an idealised lookup-table adversary, who knows exactly what network traffic each web page can produce. The adversary creates a look-up table $T(p) = \{w_1, w_2, ...\}$, that associates each packet sequence $p$ of network traffic to the set of web pages $w_i$ that can produce that traffic (Figure 2). The packet sequences of two pages are considered to be the same if all their packets have the same size, direction and time. The error of such an adversary is computed by counting the collisions (i.e., how many pages produce the same traffic), and is the smallest error achievable by any WF adversary.

**Fig. 2.** The idealized lookup-table adversary proposed by Cai et al. knows exactly what network traffic each web page ($w_1$, $w_2$, ...) may produce, and only commits an error when many pages exhibit identical traffic.

This methodology has several downsides. The smallest change in the quantities observed by the adversary (e.g., time, total bandwidth), which is likely to happen in network communications because of noise, leads similar traffic to be misclassified by the adversary. Furthermore, the accuracy of such an adversary can only be estimated for a defense if: i) the defense is deterministic, and ii) we know how it is constructed internally. The second requirement is of particular interest: if we were blindly applying this method to some defense, the results would largely underestimate both probabilistic and deterministic defenses, because of noise in network data. In fact, Cai et al. only applied this method to defenses for which, by design, only some characteristics of the traffic (e.g. the total transmission size) could change; even in this case, however, they had to apply the method to partial information (e.g., they excluded timing information), as otherwise it would have returned unreasonable bounds. We will highlight this via experiments (section 9).

Recent work proposed an extension of the method by Cai et al. for probabilistic defenses [28]. The method attempts to compute the probability distribution of network traffic defended by a defense, and to derive from it the smallest error achievable. However, to estimate such distribution i) they use a look-up table strategy, which is highly susceptible to noise, and ii) they compute it by running the defense a finite number of times. Because of these issues, the estimated distribution is not guaranteed to approximate the real distribution, and the method does not guarantee a valid bound.

## 4 Formalization of WF attacks

### 4.1 General concepts

$\mathcal{W}$ is a set containing all possible web pages. An adversary selects a subset of such pages, $\mathcal{W}_m$, which we call monitored web pages. We define a set of unmonitored web pages, $\mathcal{W}_u$, such that $\mathcal{W}_m \cap \mathcal{W}_u = \varnothing$ and $\mathcal{W}_m \cup \mathcal{W}_u \subseteq \mathcal{W}$. The victim can visit any page in $\mathcal{W}_m \cup \mathcal{W}_u$. It is natural to assume that $|\mathcal{W}_m| > 0$, and $|\mathcal{W}_m \cup \mathcal{W}_u| > 1$. We talk about *Closed World* scenario when $\mathcal{W}_u = \varnothing$. *Open World* scenario is when $|\mathcal{W}_u| > 0$. We also define the *One VS All* scenario, a subclass of Open World, where the adversary only monitors one web page: $|\mathcal{W}_m| = 1$ and $|\mathcal{W}_u| > 0$.

A packet sequence $p \in \mathcal{P}$ is a finite list of arrival time $t_j$, size $s_j$, and direction $d_j$ of packets:

$$p = ((t_j, s_j, d_j)) \quad \text{for } j = 1, 2, ... \quad ;$$

specifically, $t_1 = 0$ and $t_{j+1} \geq t_j$, $s_j \in (0, MTU]$, where $MTU$ (maximum transmission unit) is 1500, and $d_j \in \{\uparrow, \downarrow\}$. As previous work [5], we assume that a WF adversary, when observing the encrypted network traffic corresponding to a page load, gets a packet sequence as the only information.

A label $y \in \mathcal{Y}$ is associated with a packet sequence $p$ according to the web page $w \in \mathcal{W}$ that produced it.

Specifically,

$$y(w) = \begin{cases} w & \text{if } w \in \mathcal{W}_m \\ \odot & \text{otherwise} \end{cases} \quad ;$$

in other words, symbol $\odot$ represents a page that is not monitored.

A WF defense is a (possibly randomized) algorithm $D : \mathcal{P} \mapsto \mathcal{P}$ that takes as input a packet sequence, and returns a defended packet sequence. Note that $D$ also includes the trivial "no-defense" algorithm, which simply outputs its input packet sequence.

A WF adversary is a pair $\mathcal{A} = (\Phi, \mathrm{T})$ of a feature set $\Phi$ and an ML training algorithm T, as we shall now explain into details. A feature set is a list of algorithms[2]:

$$\Phi = (\phi_1, \phi_2, ..., \phi_Q) \quad ,$$

with $\phi_q : \mathcal{P} \mapsto \mathbb{R}^{d_q}$ and $d_q > 0$ for $q = 1, 2, ..., Q$; each $\phi_q$ gets as input a packet sequence, and returns a vector of $d_q$ real values. With a slight abuse of notation, we will use $\Phi$ to refer to the algorithm:

$$\Phi : \mathcal{P} \mapsto \mathcal{X} \quad ,$$

which runs each $\phi_q$ on an input packet sequence $p$, and concatenates the resulting vectors of values:

$$\Phi(p) = (\phi_1(p), \phi_2(p), ..., \phi_Q(p)) \quad ;$$

We call $\mathcal{X} = \mathbb{R}^d$ the object space, where $d = \sum_{q=1}^{Q} d_q$, and we call $x = \Phi(p)$ an object. An ML training algorithm $\mathrm{T} : (\mathcal{X} \times \mathcal{Y})^* \mapsto \mathcal{F}$ is an algorithm that accepts an arbitrary number of pairs of objects and respective labels, and returns an algorithm $f \in \mathcal{F}$ from the collection of algorithms $\mathcal{F} = \{f \mid f : \mathcal{X} \mapsto \mathcal{Y}\}$. We call $f$ a classifier. A training algorithm T is class-specific, in the sense that it returns classifiers of a specific class (e.g., logistic regression, SVM).

## 4.2 WF attacks

We shall now describe a WF attack. Consider the case in which a victim uses defense $D$ to protect her traffic. In the training phase, a WF adversary $\mathcal{A} = (\Phi, \mathrm{T})$ generates $n$ pairs of packet sequences, defended using defense $D$, and respective labels:

$$((p'_i, y_i)) = ((D(p_i), y_i)) \quad i = 1, 2, ..., n \quad ,$$

with $y_i \in \mathcal{W}_m$. The adversary first extracts objects from the packet sequences, obtaining what is commonly called the training set:

$$Z_{train} = ((x_i, y_i)) = ((\Phi(p'_i), y_i)) \quad i = 1, 2, ..., n \quad .$$

Then, he trains a classifier on the training set: $f = \mathrm{T}(Z_{train})$. In the attack phase, the victim visits a web page with label $y_{n+1} \in \mathcal{Y}$, using defense $D$, which produces a defended packet sequence $p'_{n+1}$. The adversary eavesdrops this packet sequence, extracts a test object $x_{n+1} = \Phi(p'_{n+1})$, and outputs a prediction $\hat{y}_{n+1} = f(x_{n+1})$.

The adversary is evaluated in terms of the probability that classifier $f$ misclassifies the test object $x_{n+1}$:

$$R^{\mathcal{A}} = P(f(x_{n+1}) \neq y_{n+1}) \quad .$$

In practice, this probability is estimated using the empirical error $\hat{R}^{\mathcal{A}}$, by averaging the error committed by the adversary over a test set of $n_{test}$ pairs:

$$Z_{test} = ((x_i, y_i)) \quad i = n+1, n+2, ..., n+n_{test} \quad ,$$

as follows:

$$\hat{R}^{\mathcal{A}} = \frac{1}{n_{test}} \sum_{i=n+1}^{n+n_{test}} I(f(x_i) \neq y_i) \quad ,$$

where $I$ is the indicator function.

## 4.3 Assumptions

The following analysis makes the standard i.i.d. assumption on $((x_1, y_1), (x_2, y_2), ..., (x_n, y_n), (x_{n+1}, y_{n+1}))$; that is, examples are independently sampled from the same (arbitrary) distribution. In other words, we expect that the traffic produced by each web page follows the same distribution. Violations of this assumption (e.g., the adversary trains on a different version of the web page) would be disadvantageous for an adversary [14]. No assumption is made about the underlying distribution.

We assume labels $y \in \mathcal{Y}$ are equally likely. While this assumption is not necessary, it was shown that the ratio of an adversary's success and random guessing probability, which is inversely related to the privacy parameter $\varepsilon$ we will introduce in the next section, is maximized by uniform priors [3]; similarly, $\varepsilon$ is minimized by uniform priors. Nevertheless, when true priors are known, one can use them to compute the bounds we will present.

---

[2] There is often confusion in what "features" indicate. In informal discussion, the terms are used ambiguously for both the algorithms that extract values, and the values themselves. From now on, we use "features" and "feature set" to indicate the algorithms extracting feature values. As indicated shortly, we will call "object" a vector of values extracted using features.

# 5 Bounds on WF adversaries

We present a method to provide security bounds for a WF defense, with respect to the class of WF adversaries using feature set $\Phi$:

$$\mathcal{A}_\Phi = \{\mathcal{A} \mid \mathcal{A} = (\Phi, \cdot)\} \quad ;$$

this class captures any choice of training algorithm T. The method is based on the optimality of the Bayes classifier, and on the empirical estimation of its probability of error. We then define a privacy metric based on this result, which can be used for evaluating the security of WF defenses.

## 5.1 Bayes error: an intuition

We assume that objects belonging to each label are drawn from some (unknown) probability distribution (i.i.d. assumption, subsection 4.3). For illustration sake, let us consider objects with only one feature: total communication time; this feature has indeed good discriminating power [11], and is difficult to hide using countermeasures. Figure 3 shows probability distribution estimates (KDE fits[3]) of the total communication time when loading two web pages from the WCN+ dataset; this dataset will be introduced in section 6. The goal of a classifier is to commit the smallest error at distinguishing between the two distributions. Assuming that i) the distributions shown in Figure 3 were the "true" ones, and that ii) they were known to a classifier, then the best strategy for such classifier would be predicting the most likely web page given an observation, according to the Bayes rule on these distributions. For example, if the observed communication time was 5, this classifier would predict *Web Page 1*. Under these idealized circumstances, the smallest error achievable is denoted by the uncertainty area in red: this area is called the Bayes error. Note that this performance limit derives from the distribution of data, and no classifier (even if computationally unbounded) can do better on this feature.

In practice, we typically do not know the true underlying probability distributions. However, there exist methods for estimating the Bayes error from data, without any further assumption on the distributions themselves. The remainder of this section will formalize the optimality of the Bayes error, indicate a lower bound estimate of it, and apply these results to evaluate the security of WF defenses.

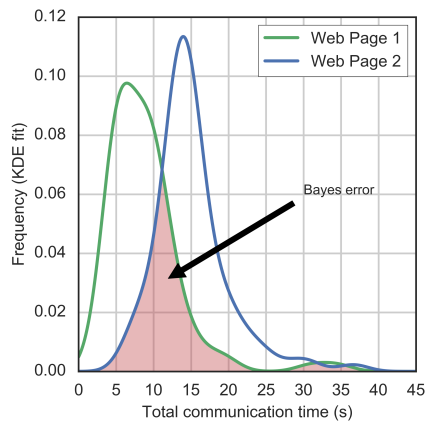

**Fig. 3.** Distributions (KDE fits) of total communication time of the first two web pages of the WCN+ dataset. The uncertainty area ("Bayes error", if the KDE fits were the true distributions) is the smallest achievable error when only using total communication time as a feature.

## 5.2 Optimality of the Bayes classifier

Let $((x_1, y_1), (x_2, y_2), ..., (x_n, y_n), (x_{n+1}, y_{n+1}))$ be i.i.d. pairs of objects $x_i \in \mathcal{X}$ and respective labels $y_i \in \mathcal{Y}$. The pairs are generated according to some distribution over $\mathcal{X} \times \mathcal{Y}$. We call training set the pairs:

$$((x_i, y_i)) \quad i = 1, 2, ..., n \quad ,$$

and we refer to $x_{n+1}$ as the test object. An ML classification task requires finding a classifier $f : \mathcal{X} \mapsto \mathcal{Y}$ which minimizes the probability of error:

$$P(f(x_{n+1}) \neq y_{n+1}) \quad .$$

For any classification task, the Bayes classifier is the idealized classifier that minimizes this error. Formally, let $\mathcal{F}$ be the collection of all classifiers $\{f \mid f : \mathcal{X} \mapsto \mathcal{Y}\}$, let $R^f$ and $R^*$ be respectively the probability of error of a classifier $f \in \mathcal{F}$ and of the Bayes classifier in predicting a label for the test object. Then:

$$R^* = \min_{f \in \mathcal{F}} R^f \quad . \tag{1}$$

A proof of this can be found in the second chapter of the book "Pattern Classification" by Duda et al. [10].

---

[3] KDE, similarly to histograms, is a non-parametric technique for estimating probability distributions; we used it here as a purely illustrative mean to introduce the concept of Bayes error. These should not be expected to be the true distributions.

The Bayes classifier is a function of prior probabilities and probability density functions over $\mathcal{X} \times \mathcal{Y}$; unfortunately, these quantities are most often unknown in practice, and the Bayes classifier is simply an ideal optimum classifier. Nevertheless, as shown in the next section, it is possible to estimate the Bayes error $R^*$ on the basis of the error of other classifiers.

### 5.3 Lower bound estimate of Bayes error

The true Bayes error is typically not known in practice, for the same reason that the Bayes classifier cannot be used (i.e., true probability distributions are unknown). Many methods, however, exist to estimate the Bayes error $R^*$. We used a well-known lower bound estimate of $R^*$, due to Cover and Hart [8], which is based on the error of the Nearest Neighbor (NN) classifier.

The NN classifier works as follows. Let $((x_1, y_1), (x_2, y_2), ..., (x_n, y_n))$ be a training set, and $x_{n+1}$ be a test object for which we want to predict the label. NN predicts for $x_{n+1}$ the label of its closest object. Closeness is determined with respect to some distance metric; the guarantees we are about to formulate are however independent of the metric. We require that $x_i$, for $i = 1, 2, ..., n+1$, take values from a separable[4] metric space $\mathcal{X}$.

The following result holds for the NN classifier. Let us consider a specific $L$-labels classification setting; i.e., $L = |\mathcal{Y}|$. Let $R^{NN}$ be the probability that the NN classifier trained on a training set of size $n$ misclassifies a test object. Then, asymptotically, as $n \to \infty$ [8]:

$$R^* \leq R^{NN} \leq R^* \left(2 - \frac{L}{L-1} R^*\right) \quad .$$

This holds for arbitrary distributions on $\mathcal{X} \times \mathcal{Y}$.

From this inequality, it is possible to derive a lower bound of $R^*$:

$$\frac{L-1}{L} \left(1 - \sqrt{1 - \frac{L}{L-1} R^{NN}}\right) \leq R^* \quad . \quad (2)$$

We can now define a lower bound estimate of the Bayes classifier's error as:

$$\hat{R}^* = \frac{L-1}{L} \left(1 - \sqrt{1 - \frac{L}{L-1} \hat{R}^{NN}}\right) \quad , \quad (3)$$

where $\hat{R}^{NN}$ is the error of the NN classifier computed on a dataset[5].

### 5.4 Bounding $\mathcal{A}_\Phi$ adversaries

From the previous discussion it follows that the error of a WF adversary using feature set $\Phi$ is asymptotically lower-bounded by $\hat{R}^*$ estimated for the same features.

**Theorem 1.** *Fix a feature set $\Phi$. Sample $Z = ((x_1, y_1), (x_2, y_2), ..., (x_{n+1}, y_{n+1}))$ i.i.d. from an arbitrary distribution, where $x_i$ take values from a separable metric space $\mathcal{X}$ and $y_i \in \mathcal{Y}$ for $i = 1, 2, ..., n+1$, let $Z_{train} = ((x_1, y_1), (x_2, y_2), ..., (x_n, y_n))$, and $L = |\mathcal{Y}|$.*

*For any training algorithm $\mathrm{T} \in \mathcal{T}$, where $\mathcal{T} = \{\mathrm{T} \mid \mathrm{T} : (\mathcal{X} \times \mathcal{Y})^* \mapsto \mathcal{F}\}$ with $\mathcal{F} = \{f \mid f : \mathcal{X} \mapsto \mathcal{Y}\}$, let $\mathcal{A} = (\Phi, \mathrm{T})$, let $R^\mathcal{A}$ be the probability that the classifier $f = \mathrm{T}(Z_{train})$ misclassifies a test object $x_{n+1}$; also, let $\hat{R}^{NN}$ be the probability of error of the NN classifier trained on $Z_{train}$ in predicting $x_{n+1}$, and let $\hat{R}^*$ be a Bayes error lower bound computed from $\hat{R}^{NN}$ and $L$ as in Equation 3. Then, as $n \to \infty$:*

$$\hat{R}^* \leq R^* \leq R^\mathcal{A} \quad ,$$

*where $R^*$ is the error of the Bayes classifier, which has perfect knowledge of the prior probabilities and probability distributions from which $Z$ was sampled.*

*Proof.* The second inequality, $R^* \leq R^\mathcal{A}$, follows trivially from the fact that the Bayes classifier commits the smallest error among the collection of classifiers $\mathcal{F} = \{f \mid \mathcal{X} \mapsto \mathcal{Y}\}$ (Equation 1), and any training algorithm $\mathrm{T} \in \mathcal{T}$ returns a model $f \in \mathcal{F}$.

Because $\hat{R}^*$ is computed as in Equation 3, for which holds Equation 2, $\hat{R}^* \leq R^*$ holds asymptotically in $n$ [8]. □

In practice, we do not have infinite examples, and we need to compute the lower bound $\hat{R}^*$ on a finite dataset. In section 7 we show empirically that the error bound estimate $\hat{R}^*$, computed for a certain set of features, is always smaller than the respective attack error.

The guarantee is independent of the NN classifier distance metric. In subsection 7.1 we compared different metrics, and selected Euclidean for experiments.

---

**4** Separability of a metric space means it contains a countable dense subset; informally, it contains a countable subset whose points are arbitrarily close to its points. Note that the Euclidean $\mathbb{R}^d$ metric space and all finite metric spaces are separable.

**5** An implementation note: $\frac{L}{L-1} \hat{R}^{NN}$ is usually smaller or equal to 1. However, because of noise in experiments, this quantity may assume a value larger than 1, making the value under square root negative. In practice, we use $1 - \min\left(\frac{L}{L-1} \hat{R}^{NN}, 1\right)$ under square root to avoid this.

## 5.5 Privacy metric

An error lower bound $\hat{R}^*$ alone can be uninformative of the setting under which it was obtained. For instance, an error rate of 50% in a binary classification problem is much worse than the same error achieved in 100-labels classification; trivially, we can obtain 50% in binary classification by uniform random guessing, if labels are equally likely, while the same strategy for 100 labels would commit 99% error on average.

We borrow from cryptography the concept of advantage ($Adv$), and use it to define a privacy metric for WF defenses. Advantage indicates how better than random guessing an adversary can do; it is typically defined in a binary classification setting, with respect to the error $R$ of an adversary:

$$Adv = 2 \left| \frac{1}{2} - R \right| \quad ,$$

where $\frac{1}{2}$ is the random guessing probability of error, and 2 is a normalizing factor. In an $L$-labels classification setting, with $L > 1$ and assuming labels are equally likely[6], random guessing error becomes $R^G = \frac{L-1}{L}$, and the generalized advantage is[7]:

$$Adv = \frac{1}{R^G} \left| R^G - R \right| \quad .$$

We introduce a privacy metric, $(\varepsilon, \Phi)$-*privacy*, for a feature set $\Phi$ and a privacy parameter $\varepsilon$ defined as:

$$\varepsilon = 1 - Adv \quad ,$$

where $Adv$ is computed in terms of $\hat{R}^*$ estimated using feature set $\Phi$. The privacy parameter takes value 1 for a perfectly private defense, where an adversary is forced into random guessing; it takes value 0 if there exists an adversary achieving an error close or equal to 0.

By reasonably assuming $\hat{R}^* \leq R^G$, the definition of $\varepsilon$ simplifies to:

$$\varepsilon = \frac{\hat{R}^*}{R^G} \quad . \quad (4)$$

For a feature set $\Phi$ and privacy parameter $\varepsilon$, we say a defense is $(\varepsilon, \Phi)$-*private* if the privacy parameter computed for $\Phi$ is equal to $\varepsilon$.

---

[6] As suggested in subsection 4.3, it is not necessary to assume that labels are equally likely. If priors on labels were known, $R^G$ would be the error of the majority classifier, $1 - \max_{y \in Y} P(y)$.
[7] Typically in cryptography the advantage is defined in terms of the success probability, rather than the error probability. The expression we provide is equivalent, and the advantage is appropriately normalized to take values in [0,1].

## 5.6 Independence from features

A natural question raises from our discussion: is it possible to have a lower bound guarantee (and thus, a privacy metric) that is independent from features?

The following result would seem to help [9]:

**Remark 1.** *Let $\mathcal{P}$ be the original data space (packet sequence space, in our setting). Let $R^*_{\mathcal{P}}$ be the Bayes error in the original space. For any transformation $\Phi : \mathcal{P} \mapsto \mathcal{X}$, mapping objects from the original space into a new space $\mathcal{X}$:*

$$R^*_{\mathcal{P}} \leq R^*_{\Phi(\mathcal{P})} \quad . \quad (5)$$

In other words, no transformation of the original data improves the Bayes error, and an estimate computed on full information would be valid regardless the choice of features.

However, in practice, if the original data is noisy and comes from a very large space, an error estimate in the original space converges too slowly to its optimum, and it is necessary to operate on transformations of the data. The same applies to the WF setting: the error bound estimate computed in the packet sequence space converges too slowly (subsection 7.4), which is also why WF researchers typically constructed attacks on feature sets and not directly on packet sequences.

In section 10, we introduce the concept of *efficient* feature set, as a future research direction towards independence from features.

# 6 Experimental Methodology

Experiments aimed at verifying the validity of the Bayes error estimate $\hat{R}^*$, evaluating improvements in features, and measuring the privacy of WF defenses using $(\varepsilon, \Phi)$-*privacy*. We conducted experiments for the most influential attacks and defenses on the dataset from the USENIX 2014 paper by Wang et al. [26][8]. Henceforth, we refer to this dataset as WCN+.

**Dataset**
The WCN+ dataset is a collection of network traces produced by visiting, over Tor, 100 unique web pages 90 times each. Since the data represents Tor traffic, where packets have a fixed size, attacks using size of packets as a feature may be penalized.

---

[8] https://cs.uwaterloo.ca/~t55wang/.

**Background noise**

The privacy bounds of a WF defense should be determined under optimal conditions for the adversary. Any noise in the data, such as that introduced by background web pages or by external applications (e.g. BitTorrent client), would produce an overestimate of the defense's privacy. For this reason, we did not add any further noise to the dataset.

**Distribution of the labels in the test set**

We designed the experiments so that training and test sets would have a uniform distribution of labels. This prevents an unbalance towards web pages that are easier to defend, which would cause overestimates of the privacy.

**Scenarios**

We performed experiments in the Closed World and One VS All scenarios. The former is of common use when evaluating defenses, and it indicates the ability of an adversary to distinguish many pages in nearly-perfect conditions. One VS All is an extreme case of Open World scenario, in which an adversary only targets one web page. This scenario highlights how well a defense protects unique pages. We remark that our method is applicable to any other case of Open World.

**Code**

The code of the major WF attacks and defenses is openly available, and it is usually written in `Python` or `C`. We acknowledge the use of the following code:
- T. Wang: Decoy Pages, BuFLO, Tamaraw, k-NN[9]
- K. P. Dyer: VNG++, LL[10]
- J. Hayes: k-FP[11]
- M. Juarez: WTF-PAD[12].

We created a `Python` framework to abstract the different APIs employed by researchers, with the goal of unifying the way experiments are carried out in WF research. We implemented CS-BuFLO, and the CUMUL attack, and adapted other researchers' code to the API. We created routines to compute error bounds and $(\varepsilon, \Phi)$-*privacy* of WF defenses as they are described in this paper.

| Distance | $\hat{R}^*$ (%) | Feature Set |
|---|---|---|
| Euclidean | 5.8 | k-NN |
| Standardized Euclidean | 6.0 | CUMUL |
| City Block | 6.6 | k-NN |
| Levenshtein | 12.2 | packets' sizes |

**Table 3.** Comparison of the Nearest Neighbor-based lower bound $\hat{R}^*$ obtained using different distance metrics. Experiments were performed against No Defense (RP), on the `WCN+` dataset. In this table, the name of an attack refers to its feature set.

## 7 Experimental Analysis of the Method

In this section, we verify the validity of the lower bound of error $\hat{R}^*$ defined in section 5 for the major attacks. We then discuss whether a privacy metric dependent on features is reasonable, by showing that feature sets from recent attacks have not improved significantly (if at all).

### 7.1 What Distance Metric

The lower bound estimate $\hat{R}^*$ depends on a distance metric. Whilst its guarantee holds regardless this choice, different metrics may perform better on our data.

We experimented with four distance metrics: Euclidean, Standardized Euclidean, City Block and Levenshtein. The first three distances were applied to feature sets; the bound of the best performing feature set is shown. We applied Levenshtein distance directly to packet sequences as follows. Given a packet sequence $((t_j, s_j, d_j))$, we created a binary string containing 0 when $d_j = \uparrow$, and 1 when $d_j = \downarrow$; this makes sense in the context of Tor traffic, where packets' sizes are the same; by doing so we excluded timing information. We then computed the NN-based bound on the binary strings using Levenshtein as a distance metric.

Results in Table 3 indicate a choice between Euclidean, Standardized Euclidean and City Block does not impact heavily on the bound. Levenshtein distance, however, performed poorly; this may be due to the fact

---

**9** https://cs.uwaterloo.ca/~t55wang/wf.html.
**10** https://github.com/kpdyer/website-fingerprinting.
**11** https://github.com/jhayes14/k-FP.
**12** https://github.com/wtfpad/wtfpad.

that it does not consider timing information. In the following experiments we used Euclidean distance.

## 7.2 Empirical validity of the bound

Under the sole i.i.d. assumption, it is not possible to prove convergence rates of a Bayes error estimate (subsection 10.3). We studied the asymptotic guarantee of $\hat{R}^*$ in finite sample conditions experimentally, for an increasing number $n$ of training examples. We used the following heuristics to verify the estimates' convergence:
– verify that $\hat{R}^*$ computed for a feature set is lower than the error of an attack performed with the same feature set;
– visually inspect that the decreasing trend of $\hat{R}^*$ becomes slower as $n$ increases.

We performed these experiments in a Closed World scenario on the WCN+ dataset, which is composed of $n_{tot} = 9000$ network traces from 100 distinct pages. We kept a fixed test set of size $n_{test} = 0.2 \times n_{tot}$, and iteratively increased the number of traces of the training set: $n = \{0.1 \times n_{tot}, 0.2 \times n_{tot}, ..., 0.8 \times n_{tot}\}$. To show that the bound is valid for adversaries predicting new data, we computed the bound only on the training set, and compared it with the attack error on the test set. More specifically, for an adversary $\mathcal{A} = (\Phi, T)$, we computed the attack error $R^{\mathcal{A}}$ by training the attack's classifier on the training set, and predicting the test set. We computed the lower bound $\hat{R}^*$, for the same feature set $\Phi$, on the training set, using 5-folds cross validation (CV) to reduce the variance.

Figure 4 presents the results of these experiments. Each color represents a set of features; continuous lines indicate the error of an attack, and dashed lines error bounds. The figure shows that, for any size of the training set, the bound estimate computed for a feature set is a lower bound of the error committed by an attack using the same features. Furthermore, the decreasing trend of the estimates seems to converge towards an asymptote.

## 7.3 Evolution of feature sets

The Bayes error indicates the smallest error achievable when using a certain feature set. We use $\hat{R}^*$ to compare the feature sets of major WF attacks, and address the question: did feature sets improve in recent attacks?

We computed the bounds in a Closed World scenario, on the full WCN+ dataset, using 5-folds CV. Table 4 shows the evolution of feature sets in the major attacks from 2006 to 2017. We observe that, against most defenses, the feature set used by the 2014 attack k-NN [26] performs nearly as well as the one employed by the current state-of-the-art attack, k-FP [12].

These results support the following hypothesis: improving feature sets is becoming harder, and there may be a limit for their improvement. This strengthens the value of a privacy metric depending on a feature set.

We remark however that these findings should not discourage WF studies to improve classifiers: while $R^*$ is the smallest error achievable by a classifier, *a priori* we do not know which classifier will reach it. Furthermore, as suggested by Devroye et al. (chapter 32 in reference [9]), if the goal is improving an attack, then features should be optimized with respect to the chosen classifier's empirical error rather than $\hat{R}^*$; the reason they give is that no Bayes error estimate can guarantee good performance (subsection 10.3).

## 7.4 $\hat{R}^*$ on full information

Instead of selecting a set of features, we estimated the error bound $\hat{R}^*$ using full information, and compared it with the ones achieved by the best feature sets.

We concatenated time and size of each packet sequence to form a vector, and padded such vector to the same length. We used Euclidean distance in this space to estimate the Bayes error.

The bounds obtained on full information were worse than those achieved using feature sets (Figure 5). This result is not surprising: as we discussed in subsection 5.6, in the real world, classifiers tend to converge more rapidly in the feature space than in the original space. This shows that in the context of WF we need to rely on feature sets. However, we notice that, for some defenses (Tamaraw, CS-BuFLO), the bounds computed on full information are close to the error of the best classifiers (Appendix B). This suggests their bounds may be valid independently of the chosen feature set.

Future research may also find a better performing distance metric, which might favor a quicker convergence of the Bayes error estimate directly computed in the packet sequence space.

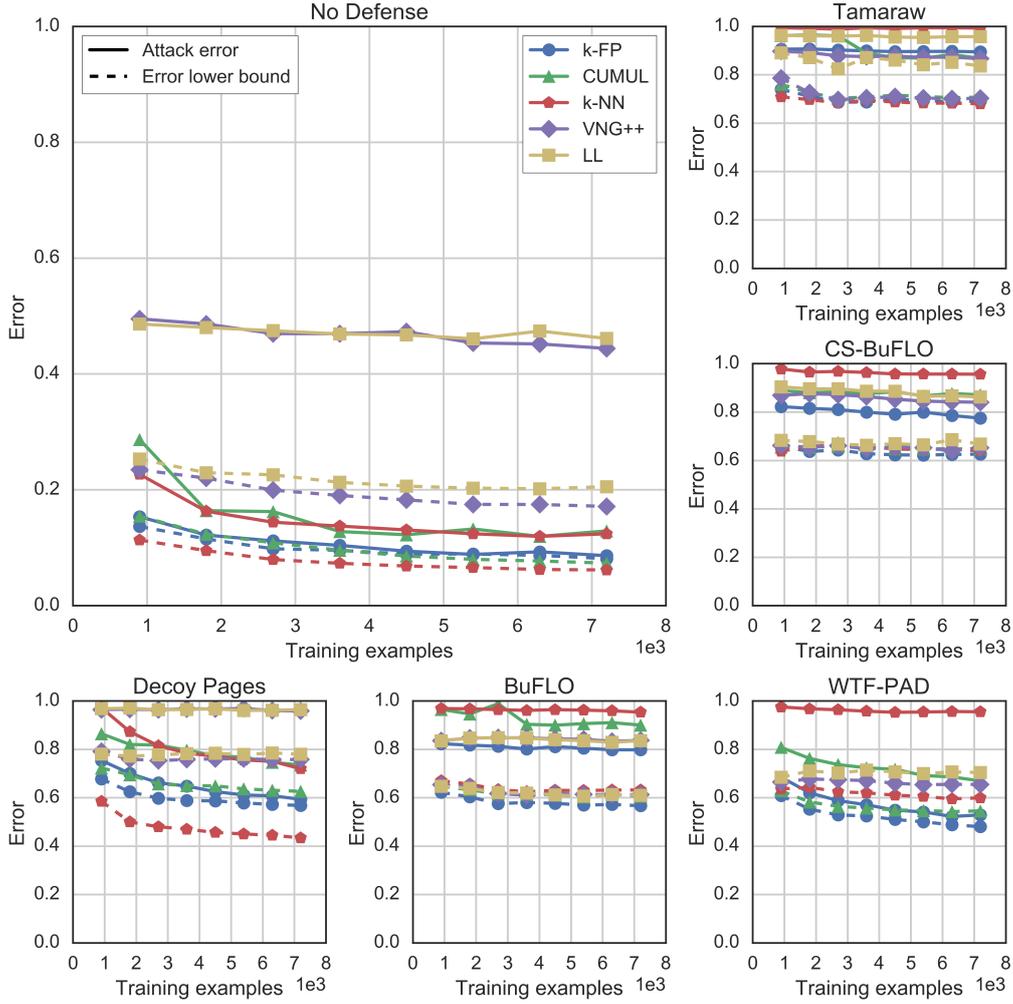

**Fig. 4.** Validity of the lower bound $\hat{R}^*$ on the WCN+ dataset (Closed World scenario), with respect to an increasing number of training examples. The lower bound estimates $\hat{R}^*$ (dashed lines) are smaller than the error of an attack (continuous lines) performed using the same feature set (same color).

## 8 Evaluation of Defenses

We measured $(\varepsilon, \Phi)$-*privacy* and overheads of the defenses. We computed error bounds and $(\varepsilon, \Phi)$-*privacy* under Closed World and One VS All scenarios. For this experiment, we used the full WCN+ dataset.

We computed the packet overhead $Oh_D^b$ for each packet sequence $p$ and a defense $D$ as:

$$Oh_D^b = \left(\frac{|D(p)|}{|p|} - 1\right) \times 100 \quad,$$

where $|p|$ indicates the number of packets in packet sequence $p$. We computed the time overhead $Oh_D^t$ by dividing the difference $t_\ell - t_1$ for packets in packet sequence $p$ by the same computed for packets in $p'$. We report the median overhead among packet sequences for both $Oh^b$ and $Oh^t$.

For the Closed World scenario, we considered all the dataset, extracted the features, and computed $\hat{R}^*$ (Equation 3) using 5-folds CV. We then determined $(\varepsilon, \Phi)$-*privacy*. Random guessing error in this scenario (100-labels classification setting) was 0.99.

In the One VS All case, for each potential target $w \in \mathcal{W}$, we took all 90 examples from web page $w$, and then randomly selected 90 examples from other pages, which we labeled $\odot$. Finally, we computed the bounds in this binary classification setting, where random guessing error was 0.5. Since the method produced a bound for each page, indicating its fingerprintability, we averaged these results. In this scenario, we observed a high variance, which is due to the heterogeneity of the pages.

|  | Liberatore et al. (2006) | Dyer et al. (2012) | Wang et al. (2014) | Panchenko et al. (2016) | Hayes et al. (2016) |
|---|---|---|---|---|---|
| No Defense | 20.8 | 17.2 | **6.2** | 7.3 | 8.2 |
| Decoy Pages | 77.8 | 74.9 | **42.6** | 62.5 | 56.2 |
| BuFLO | 60.6 | 60.9 | 62.8 | 60.0 | **56.9** |
| Tamaraw | 84.4 | 69.8 | **69.0** | 70.3 | **69.0** |
| CS-BuFLO | 66.4 | 64.5 | 62.9 | 63.0 | **61.9** |
| WTF-PAD | 69.8 | 65.3 | 60.2 | 54.9 | **48.6** |

**Table 4.** *Evolution of feature sets.* The smallest error achievable ($\hat{R}^*$) when using the feature sets of the major attacks up to 2017, against the most influential WF defenses. The bound of the best feature set for each defense is in bold. Feature sets have not improved substantially since the attack by Wang et al. (2014). Bounds were computed on the WCN+ dataset in a Closed World scenario.

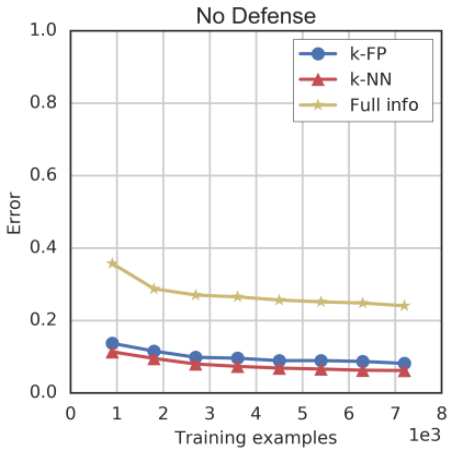

**Fig. 5.** Comparison of lower bounds $\hat{R}^*$ computed using full information (i.e., in the packet sequence space) and those computed using the feature sets of the best attacks (k-NN, k-FP). Bounds computed on full information do not converge rapidly enough, and are invalidated by those using feature sets.

We also considered taking the minimum of the bounds as an alternative to averaging. Unfortunately, to date no defense is able to achieve a minimum bound better than 2.9% (Tamaraw). Future work may consider alternative methods to combine error bounds in this scenario.

Table 1 shows the results in both scenarios. Results indicate that the best defense for the Closed World scenario is Tamaraw. However, we notice that BuFLO and Decoy Pages outperform Tamaraw in the One VS All scenario. This suggests that Decoy Pages, a very simple defense to implement, could be a safe choice for defending individual pages from WF attacks. Results also confirm that Tor Browser's default defense (RP) is ineffective, as previously suggested by attacks [6, 14, 27].

| Defense | $\hat{R}^*$ (%) | Cai et al. (%) | Cai et al. - full (%) |
|---|---|---|---|
| BuFLO | 57 | 53 | 19 |
| Tamaraw | 69 | 91 | 11 |

**Table 5.** Comparison of the bound for deterministic defenses suggested by Cai et al. and the lower bound estimate $\hat{R}^*$. The last column indicates the bound by Cai et al. on full information, which indicates its susceptibility to noise.

## 9 Comparison with previous evaluation

We computed the bound proposed by Cai et al. [5] on the full WCN+ dataset. As discussed in subsection 3.3, this method is only applicable to deterministic defenses, and it needs to be adapted to the defense.

We estimated the bound by Cai et al. for defenses BuFLO and Tamaraw. With the goal of reducing noise, the method requires limiting the information available to an adversary: against BuFLO, an adversary can only observe the transmission size; against Tamaraw, the number of incoming and outgoing packets.

Table 5 compares the bound by Cai et al. with $\hat{R}^*$. Results do not exhibit any particular trend: the bound by Cai et al. is larger than $\hat{R}^*$ for Tamaraw, but smaller for BuFLO. We notice, however, that for Tamaraw their error bound is higher than the error of the best performing attack (86.6%, VNG++) (Figure 4). This may be due to noise in the data.

To better grasp the impact of noise on their method, we estimated the bound using all the information available to a real-world adversary (Table 5). Interestingly, in this case the error bound for Tamaraw was smaller than the one for BuFLO. This indicates that the lookup-table strategy is strongly affected by noise, which suggests that it may not be reliable for evaluating security.

# 10 Discussion and Conclusion

Past WF attacks employed new combinations of ML classifiers and feature sets to defeat defenses. We showed that a Bayes error lower bound estimate $\hat{R}^*$ bounds the error of a WF adversary using a certain feature set, regardless his choice of classifier. On this basis, we introduced privacy metric $(\varepsilon, \Phi)$-*privacy* for WF defenses, based on $\hat{R}^*$, that indicates how far a defense is from a perfectly private one. We shall now discuss open questions and future directions.

## 10.1 Towards $\varepsilon$-*privacy*

Our work encourages WF research to shift its focus towards features. It further gives a tool to help future studies in this direction: the error estimate $\hat{R}^*$, a lower bound of the error of any adversary using a certain feature set, can also be used to evaluate feature sets independently of a classifier (Table 4).

Whilst our results suggest that finding better performing features is becoming harder, we recognize that one limitation of $(\varepsilon, \Phi)$-*privacy* is its dependence on a set of features. Future research may attempt to formulate a similar guarantee independently of features. We call $\varepsilon$-*privacy* such a guarantee, which is defined as follows. Let $\mathcal{P}$ be the range of a defense $D$. Let $\Phi^*$ be the set of all the features that can be extracted from $p \in \mathcal{P}$. Then $D$ is $\varepsilon$-*private* if it is $(\varepsilon, \Phi)$-*private* for some $\Phi$, and for any $\Phi' \subset \Phi^*$, guaranteeing $(\varepsilon', \Phi')$-privacy, holds:

$$\varepsilon \leq \varepsilon' + \delta \quad ,$$

for an arbitrary $\delta \geq 0$; that is, $D$ is $\varepsilon$-*private* if the value of $\varepsilon$ achieved by a feature set $\Phi$ is proven to be the smallest up to some negligible additive term $\delta$.

## 10.2 Computational bounds

Whilst recent WF research attempted to optimize the computational time and storage required to perform attacks (e.g., [12, 27]), researchers have not imposed computational bounds to WF adversaries. Further research could look at ways to improve the lower bound estimates for computationally bounded WF adversaries, instead of the unbounded adversary we modeled in this paper.

In the context of feature sets, we suggest future research should also direct its attention towards *efficient* feature sets, as we shall now define. Let $R^*_\mathcal{P}$ be the Bayes error committed when using full information. An efficient feature set $\Phi$ is a feature set that satisfies:
1. $R^*_{\Phi(\mathcal{P})} = R^*_\mathcal{P} + \delta$ for a negligible $\delta \geq 0$;
2. a Bayes error estimate on $\Phi(\mathcal{P})$ converges quickly;
3. $\Phi : \mathcal{P} \mapsto \mathcal{X}$ is bounded in time and memory.

## 10.3 Bayes error estimates

We opted for a well-studied asymptotic lower bound estimate of the Bayes error, which is based on the error of the NN classifier. Future work may explore other estimates in order to obtain tighter bounds. Unfortunately, i) no Bayes error estimate can guarantee on its performances (Theorem 8.5 in reference [9]), ii) under the relaxed i.i.d. assumption on data it is not possible to prove any convergence rate for a Bayes error estimate [1]; we will thus need to evaluate convergence experimentally for any estimate.

In our experiments we also evaluated a Bayes error lower bound estimate that is based on the error of the k-NN classifier (Appendix A). Results indicate this estimate has an undesired property: its validity highly depends on the chosen parameters. It is although possible to use such estimate with a careful choice of parameters. A further option is to use an ensemble of Bayes error estimates; for example, see reference [25].

When the true distributions are known, however, it is possible to compute the Bayes error directly. Recent probabilistic defense ALPaCA samples web pages' characteristics from known distributions [7]; it may be possible in this case to compute the Bayes error by taking into account such distributions.

## 10.4 $\hat{R}^*$ and $(\varepsilon, \Phi)$-*privacy*

Both $\hat{R}^*$ and $(\varepsilon, \Phi)$-*privacy* are valid metrics for evaluating the privacy of a defense against WF attacks. $(\varepsilon, \Phi)$-*privacy*, which is related to the concept of advantage widely used in cryptography, gives a more precise idea of how far a defense is from a perfect one (that is, one that forces an adversary into random guessing), and should thus be preferred to evaluate the security of defenses. The privacy parameter $\varepsilon$ is also related to the *multiplicative leakage*, introduced in Quantitative Information Flow, which indicates how much information a system leaks about its input [3]. $\hat{R}^*$ may be used for comparing feature sets independently of a classifier.

## 10.5 Other applications of the method

Future research may apply our method to evaluate the security against other ML-based attacks. If an adversary uses an ML classifier to violate security properties, we can immediately estimate security bounds for the proposed countermeasures in a black-box manner. For example, in the case of traffic analysis, we can directly evaluate the ability of a defense to protect the underlying behavior of any source of data. Furthermore, if the information available to the adversary in some ML-based attack is limited (e.g., data has small dimensions), the estimated bounds in such context would be immediately feature-independent, giving $\varepsilon$-*privacy* guarantees.

## 10.6 Conclusions

We showed that there exist natural bounds on any WF adversary, and that we can estimate and use them to evaluate the privacy of any WF defense in a black-box manner. We expect this will be used to evaluate the privacy of padding schemes in protocols such as Tor, SSH, and TLS.

We hope our work inspires further research to formalize security for WF and generic ML-based attacks, for even more realistic adversaries (e.g., computationally bounded adversaries). We suspect this will help moving from $(\varepsilon, \Phi)$-*privacy* to $\varepsilon$-*privacy*. Nonetheless, we reiterate that our framework already provides security guarantees, and that it allows future proposals and evaluations of WF defenses to only focus on the state-of-the-art features in order to show their privacy.

## 11 Acknowledgments

The author was supported by the EPSRC and the UK government as part of the Centre for Doctoral Training in Cyber Security at Royal Holloway, University of London (EP/K035584/1).

I thank the anonymous reviewers for their comments, that led to presentation improvements. I am grateful to Kenneth Paterson for advice on formalization and presentation. I thank Alexander Gammerman, Ilia Nouretdinov, Alex Davidson, Thomas Ristenpart, Jamie Hayes, Marc Juarez, Kostas Chatzikokolakis and Vladimir Vovk for helpful discussion. I thank Gregory Fenn, Joseph O'Brien and Joanne Woodage for revising previous versions of this paper.

## A $k_n$-NN Bayes error estimate

An optimistically biased estimate of $R^*$ is the resubstitution error $R^{(R)}$ of the $k_n$-NN classifier [9]. The resubstitution error is the error committed by a classifier on its training set. The $k_n$-NN classifier is a k-NN classifier for which the number of neighbors $k_n$ changes in function of $n$. Let $k_n$ be such that $k_n \to \infty$, and $k_n/n \to 0$ as $n \to \infty$. Then $R^{(R)}$ converges (from below, in expectation) to $R^*$ as $n \to \infty$ [24].

Convergence from below would be an interesting property for our purposes. Unfortunately, our results suggest this bound's validity strongly depends on the choice of $k_n$. We considered $k_n = \sqrt{n}$ and $k_n = \log(n)$. Figure 6 compares the $k_n$-NN bounds with the NN bound (Equation 2) and with the best attack's error. While the $k_n$-NN bound with $k_n = \log(n)$ slightly improves on the NN bound, $k_n = \sqrt{n}$ is always invalidated. For this paper, we conservatively relied on the NN bound. However, we highlight that the $k_n$-NN estimate may be used with an appropriate choice of $k_n$.

## B Lower bounds on full information

Figure 7 shows all the bounds computed on full information, as explained in subsection 7.4.

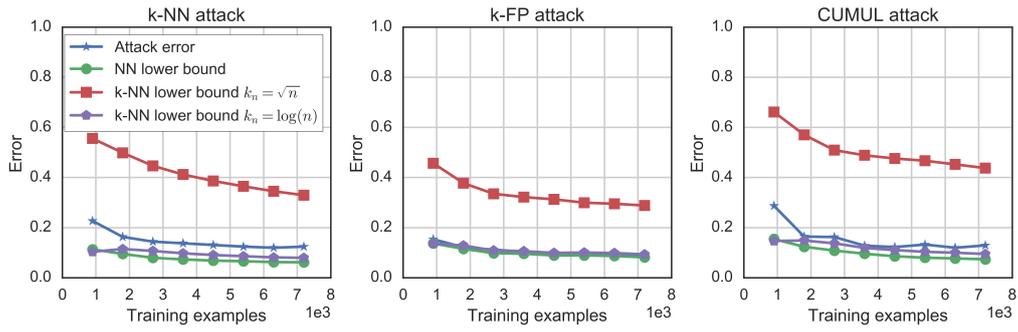

**Fig. 6.** Comparison of $k_n$-NN bounds with the NN bound $\hat{R}^*$ and attacks' error. These experiments were done against "No Defense" in a Closed World scenario.

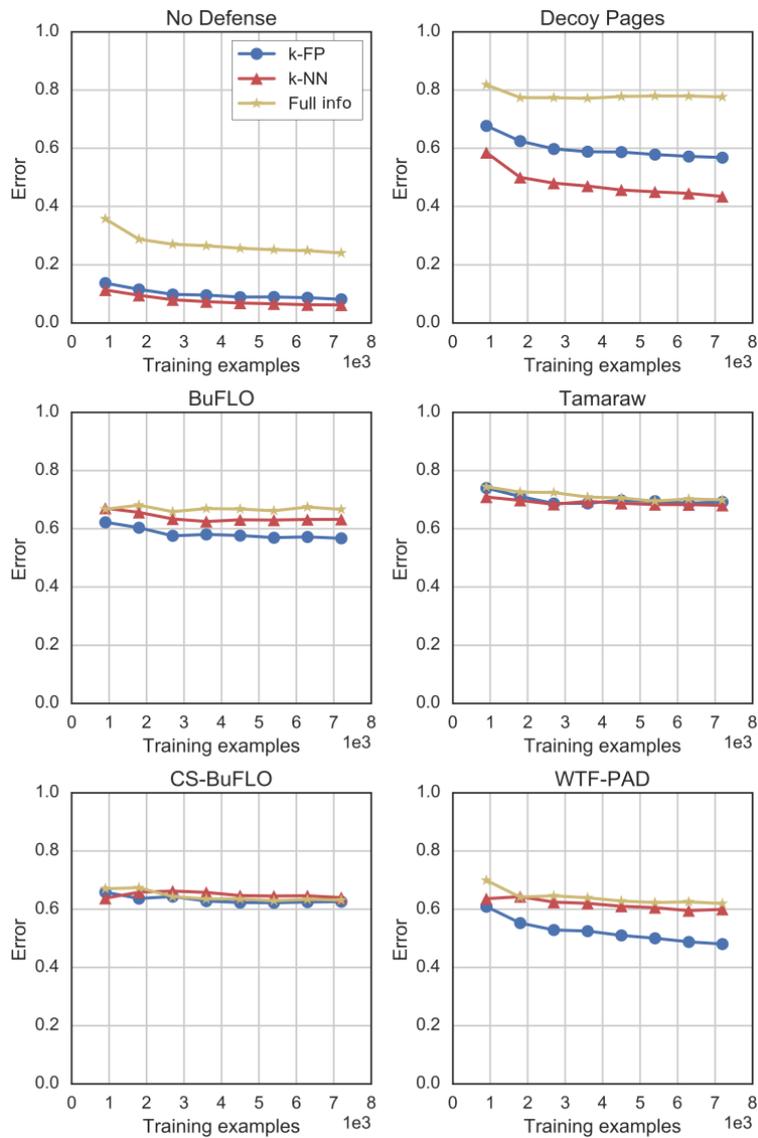

**Fig. 7.** Comparison of lower bounds $\hat{R}^*$ computed using full information (i.e., packet sequences) and those computed using the feature sets from the best attacks (k-NN, k-FP) on the WCN+ dataset. Results show that bounds using full information typically do not converge quickly enough.